\documentclass[%
reprint,
 amsmath,amssymb,
 aps,
]{revtex4-1}

\usepackage{graphicx}
\usepackage{dcolumn}
\usepackage{bm}

\begin{document}

\preprint{APS/123-QED}

\title{First order phase transition in a self-propelled particles model with 
variable angular range of interaction.}

\author{Mihir Durve} 
\email{mihirdurve@physics.unipune.ac.in}
\author{Ahmed Sayeed}
\email{Correspondence to: sayeed@physics.unipune.ac.in}
\affiliation {Department of Physics, Savitribai Phule Pune University, Pune, 
India 411007}

\date{\today}

\begin{abstract}
  We have carried out a Monte Carlo simulation of a modified version of Vicsek 
model for the motion of self-propelled particles in two dimensions. In this
model the neighborhood of interaction of a particle is a sector of the circle
with the particle at the center (rather than the whole circle as in the original
Vicsek  model). The sector is centered along the direction of the velocity of
the particle, and the half-opening angle of this sector is called
the `view-angle'. We vary the view-angle over its entire range and study the
change in the nature of the collective  motion of the particles. We find that
ordered collective motion persists down  to remarkably small view-angles. And at
a certain critical view-angle the collective  motion of the system undergoes a
first order phase transition to a disordered  state. We also find that the
reduction in the view-angle can in fact increase the order in the system
significantly. We show that the directionality of the interaction, and not only
the radial range of the interaction, plays an important role in the
determination of the nature of the above phase transition.

\end{abstract}

\pacs{Valid PACS appear here}
\maketitle


\section{Introduction}
There are many systems, natural as well as artificial, that consist of
moving and interacting `self-propelled' agents. Biological  systems such as
schools of fish~\cite{becco}, flocks of birds~\cite{couzin,ballerini}, bacterial
colonies~\cite{zhang,darton,peruani}, artificial systems such as `Kobots'
(robots specially developed for the study of flocking)~\cite{turgut}, 
platinum-silica particles in hydrogen peroxide solution~\cite{ke}, carbon coated
Janus particles in water-lutidine mixture~\cite{buttinoni},  and vibrating
rods~\cite{kudrolli,aranson,narayan} are some examples. One
remarkable characteristic of these systems is that under certain conditions
they are capable of displaying extraordinary collective dynamics, such as
highly cooperative collective motion and complex moving
patterns~\cite{vicsek,chate,tu,toner,levin,gregoire,ginelli}.

One of the simplest models proposed to describe the motion of a collection of
agents which have a tendency to `move as their neighbors do' (birds in a flock
is an obvious example) is the one by Vicsek et al. \cite{vicsek}, now commonly
known as the Vicsek model. In this model a collection of point particles move
with the same, constant speed, and each particle at discrete time intervals
`adjusts' its direction of motion so as to move along the mean direction
of motion of the particles in its local (short range) neighborhood. This
direction adjustment is imperfect due to the presence of noise in the system.
Vicsek et al. found that the nature of the collective motion of the particles
depends on the level of this noise and the particle density of the system. For
high densities and low noise levels the collective motion attains an ordered
state in which the particles move largely in a common direction. For low
densities and high noise levels there is no such collective motion and particles
essentially perform uncorrelated random walks. And as the noise level is varied
(for a fixed particle density), or as the particle density is
varied (for a fixed noise level), the system displays a non-equilibrium
order-disorder phase transition. This  subject has been recently reviewed by
Vicsek \cite{zafeiris} and Menzel \cite{menzel}. 

A number of variants of the Vicsek models have been studied, and among them
of particular interest to the present work are the ones where
certain constraints are imposed  on the  angular range of the local
neighborhood of a particle, or on the extent of the reorientation of the
particle's direction of motion. These studies have found that these
restrictions, which might naively be expected to reduce the degree of order in
the system, can in fact enhance it. Tian et al.~\cite{tian} as well as Li et 
al. \cite{li} found that restricting the angular range of interaction can reduce 
the `consensus time' (the time taken by the system to attain the stationary 
value of the order parameter). Similarly Gao et al. \cite{gao} found that 
restricting the angle of velocity reorientation can increase the order 
parameter. In the model that they studied, in a single update particle
directions are allowed to change within a limited range. And Yang et al.
\cite{yang}  found that discarding short range interactions can increase the
order parameter.

In this work we too study the effects of variation in the angular range of 
interaction on the collective motion of the particles in the Vicsek model. One
of the motivations for the Vicsek model was to understand the collective
motions of the large moving groups of living beings, such as flocks of bird or
schools of fish, where each agent is expected to move, as far as possible, in
the same direction as its close neighbors.  In the Vicsek model the
neighborhood observed by the agent is a circle (for a two dimensional system)
centered on the current position of that agent, regardless of its direction of
motion. This is quite  unrealistic, because an agent such as a bird or a fish
does not have a  $360^\circ$ view of its surroundings. For example, the
cyclopean field of view (i.e., combined field of view of both eyes
\cite{mccomb}) of the grey-headed Albatross is about $270^\circ$ in the
horizontal plane \cite{martin}, and that of \emph{Dasyatis sabina} fish is about
$327^{\circ}$ in the  horizontal plane \cite{mccomb}. This kind of restricted
view of the neighborhood also  plays a role in the phototactic motion of certain
marine organisms such as \emph{Platynereis} larvae \cite{jekely}. Thus
exploration of the effects of limitation of the angular range of interaction
neighborhood should be of interest in the study of all those processes where the
Vicsek model is relevant.

In this report we present the results of the numerical simulations of the
Vicsek model in which the angular range of the interaction neighborhood is
restricted. We have measured the order parameter of the system as a function of
the angular range of the interaction neighborhood (radius of the neighborhood
held fixed). We find that the order parameter of the system varies
non-monotonically as the angular range of the interaction neighborhood is
decreased, and at a certain point the system undergoes a first order (i.e.,
discontinuous) phase transition to a disordered state. We have also measured the
variation of the order parameter as the radius of the interaction is reduced
(without restricting the angular range of the interaction neighborhood), and
find that the resulting change in the nature of the collective motion is
qualitatively
different.

\section{Model}
\begin{figure} [!ht]
\includegraphics[trim=0cm 0cm 0cm 0cm, scale = 0.25]{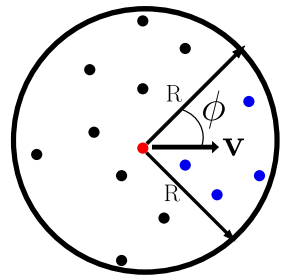}
\caption{\label{neighborhood}
(Color online) 
The neighborhood $\text{S}_i$ of the $i^{\text{th}}$ particle. The
$i^{\text{th}}$ particle is shown as the red dot, and $\text{S}_i$ is the sector
bound by the two radii marked as $R$. The red dot and the blue dots indicate
the particles lying within the neighborhood, and the black dots indicate
particles outside it.  The vector $\mathbf{v}$ indicates the direction of the
velocity of the $i^{\text{th}}$ particle. The view-angle $\phi$ is the half
opening angle of the neighborhood at the center.} 
\end{figure}

The model of self-propelled particles we have studied is a modification of the 
Vicsek model. In this model the interaction neighborhood of a particle is not 
a circle centered on that particle, but a sector of this circle, as illustrated 
in Fig. \ref{neighborhood}. The neighborhood sector $\text{S}_i$ has an
opening angle of $2\phi$ and is centered about the direction of velocity of the
$i^{\text{th}}$  particle. We shall call the half opening angle $\phi$ as the
`view-angle', which can vary from $0$ to $\pi$. For $\phi = \pi$ this model
reduces to the original Vicsek model.

Simulations are carried out in a box of size $L \times L$ with $N$ particles,
with usual periodic boundary conditions in both directions. The mean particle
density is given by $\rho = N/L^2$. The initial positions $\mathbf{r}_i$ ($i =
1, 2, 3, \ldots , N$) are assigned randomly with uniform probability within the
box. The initial directions of the particles $\theta_i$  are also assigned
randomly in the range $[-\pi, \pi]$ with uniform probability. All the particles
have the same, constant speed $v_0$.  The velocities and the positions of the
particles at time $t+1$ are obtained from  the velocities and positions at time
$t$ using the following update rules. First we update velocities of all the
particles simultaneously with
\begin{equation}
 \mathbf{v_{i}} (t+1) = v_{0} \mathcal{R}(\theta) \hat{\mathbf{v}}(t), 
\label{v_update}
\end{equation}
where $\hat{\mathbf{v}}(t)$ is the unit velocity in the direction of the mean 
velocity of the particles in the neighborhood $\text{S}_i$ of the
$i^{\text{th}}$ particle, \emph{including} the $i^{\text{th}}$ particle 
itself (see Fig. \ref{neighborhood}), and is given by
\begin{equation}
 \hat{\mathbf{v}}(t) = \frac{ \sum \limits_{j\in{ \text{S}_i}} 
{\mathbf{v}_j}(t) }
 { \left \vert  \sum \limits_{j\in{ \text{S}_i}} 
{\mathbf{v}_j}(t)  \right \vert }
 \label{vhat}.
\end{equation}
$\mathcal{R}(\theta)$ is the rotation operator which rotates 
the vector it acts upon ( i.e., $\hat{\mathbf{v}}(t)$ ) by an angle $\theta$. 
The angle $\theta$ is a random variable uniformly distributed over the interval
$[-\eta \pi, \eta \pi]$. Here $\vert \ldots \vert$ denotes the norm of the
vector, $\eta$ is the level (i.e., amplitude) of the noise that can be
varied from $0$ to $1$.

Following the velocity updates, the positions are updated with
\begin{equation}
 \mathbf{r}_i(t+1) =  \mathbf{r}_i(t) + \mathbf{v_{i}}(t+1) \Delta t,  
\label{r_update}
\end{equation}
where $\Delta t = 1$. This update scheme is known as the `forward update' in
the literature \cite{baglietto}.

To quantify the degree of order in the collective motion of the particles a 
scalar order parameter $\psi(t)$ is defined as 
\begin{equation}
 \psi(t) = \frac{1}{Nv_0} \hspace{1mm}  \hspace{1mm}\left \vert \sum_{i=1}^{N} 
\mathbf{v}_{i}(t)\hspace{1mm} \right \vert. \label{eq:orpar}
 \end{equation} 
It can be readily seen that in the perfectly ordered state when all the 
particles are moving in the same direction $\psi(t) = 1$, and in the completely 
disordered state when the directions of motion are completely random $\psi(t) =
0$ (in the limit of $N \rightarrow \infty$). In this report we use the phrase
`ordered state' to mean the stationary state of the system for which $\psi(t)
> 0$ in the limit of $N \rightarrow \infty$.
 
\section{Simulation Details}
In this work the data presented in Figs. \ref{psi_vs_phi} to \ref{hysterisis}
is produced with the following parameters fixed: the number density  $\rho =
N/L^2 = 1$, the particle speed $v_0 = 0.5$, the noise level $\eta = 0.3$, and
the interaction neighborhood radius $R = 1$.  For the data presented in Figs.
\ref{areatrans} and \ref{R_vary} the parameter $R$ is varied from $0$ to $1$,
with all other parameters same as above. Box sizes $L$ in the range of $16$ to
$36$ (i.e., particle numbers $N$ in the range $256$ to $1296$) were used.  We
have measured the order parameter $\psi(t)$ as a function of the view angle
$\phi$, which was varied over the entire range $0$ to $\pi$.  Measurements were
averaged over $20$ independent realizations, with each realization consisting of
$10^5$ to $10^7$ time-steps. In the results we have presented in the following
section we have used the time-averaged order parameter $\langle \psi(t)
\rangle$, measured as a function of the view-angle $\phi$. For satisfactory
averages we have to have a time series $\psi(t)$ of a length $T$ much
larger than the correlation time for that series. The length of the time series 
$T$ is effectively the length of a single realization multiplied by number of 
independent realizations. Near the critical value of $\phi$, the $T$ 
values are $5.2\times10^5, 4.9\times10^6, 1.3\times10^7, 1.9\times10^8$ for 
system sizes $L=16,20,24,28$ respectively. The correlation time $\tau$ can be 
estimated from the autocorrelation function for $\psi(t)$ as 

\begin{equation}
 C(\Delta t) = 
 \frac{\langle (\psi(t+\Delta t) - \overline{\psi}) (\psi(t) - 
\overline{\psi}) \rangle} {\sigma_{\psi}^2}  \label{acf},
\end{equation}

where $\overline{\psi}$ and $\sigma_{\psi}^2$ are the mean and variance of 
$\psi(t)$ for a  single time series. The angular brackets indicate averaging
over all the initial time instants $t$. The correlation time $\tau$ can be
estimated by fitting  the autocorrelation function $C(\Delta t$) to
the exponential decay function $e^{-\Delta t /\tau}$ (when that is possible),
or by using the `integrated correlation time' definition

\begin{equation}
 \tau = \sum_{\Delta t=1} C(\Delta t), \label{integtau} 
\end{equation}
where the sum is cut off at the first negative value of $C(\Delta t)$. We
have used this definition to estimate $\tau$ \cite{gould}. Near a phase
transition the correlation time $\tau$ increases rapidly with the system size
$L$, and for a given system  size it increases rapidly as the phase transition
is approached.  We  estimate that close to the transition (i.e for $|\phi -
\phi_c| \lesssim 0.01 $) $\tau \approx 4 \times 10^2, 3\times 10^3, 2.7 \times
10^4,5.5 \times 10^5$ for $L = 16, 20, 24, 28$ respectively. Thus we have the 
total length of the time series $T$ much larger than the  correlation time. For 
the results presented in the following section 
$T=13000\tau,1650\tau,510\tau,360\tau$ are for the system sizes $L=16,20,24,28$ 
respectively, close to the transition. Further away  from the
transition the $T$ values come out to be much larger multiples of $\tau$. 

\section{Results and Discussion}
   Tian et al.\cite{tian} and Li et al. \cite{li} have studied one aspect 
of the question that we are interested in the present work.  They 
varied the view angle $\phi$ and measured the `consensus time', that is, the 
time taken for the system to achieve the stationary value of the order 
parameter $\psi$, in the absence and presence of the noise.  They made a
counterintuitive observation that the consensus time can be shorter for $\phi <
\pi$, i.e., restricting the angular range of  the particle interactions can
speed up the establishment of the ordered collective motion. They found that
there is an optimum value for the view-angle for which the consensus time is the
shortest. A similar observation was made by Wang et al. \cite{wang}, who
considered not only variable view-angle  but also interaction strengths weighted
by the separations between the particle and its neighbors.

With the above mentioned earlier studies in mind, we were interested in the
effect of varying the view angle $\phi$ on the nature of collective motion of
the particles in the  modified Vicsek model considered in this study. We have
made three important observations: (i) It is not only that the collective motion
attains consensus  faster with restricted view angle (as found in the reports
\cite{tian, li}), but the stationary value of order parameter can also increase
with the decreasing view angle. (ii) The ordered state can persist down to quite
low view-angles for non-zero noise (iii) As the view angle is reduced, the
system undergoes a first order phase transition from the ordered state to the
disordered state.
  
\begin{figure} [!h]
\includegraphics[trim=0cm 0cm 0cm 0cm, scale = 0.5]{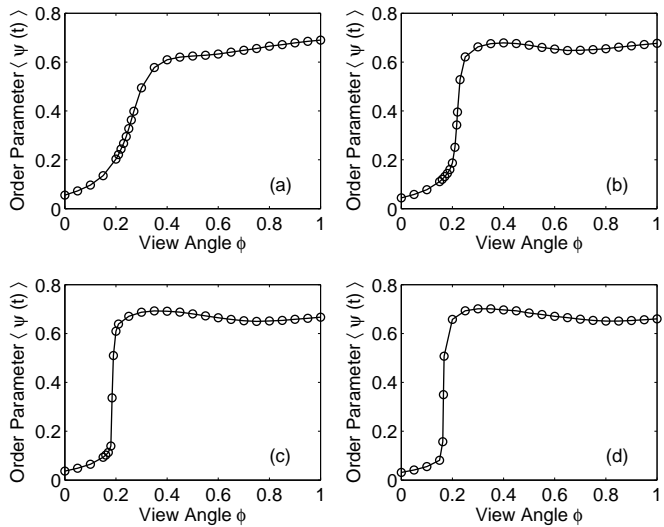}
\caption{\label{psi_vs_phi}
Order parameter $\langle \psi(t) \rangle$ \emph{vs}  view-angle $\phi$ (in units
of $\pi$) plots
for system sizes (a) $L=16$ (b) $L=20$ (c) $L=24$ (d) $L=28$.
}
\end{figure}

  In Fig. \ref{psi_vs_phi} we have shown the time-averaged order parameter 
$\langle \psi(t) \rangle$  as a function of the view angle $\phi$ for four 
different system sizes. (In this and all other plots the symbols are the data
and the connecting lines are guides to the eye.) The phase transition is clearly
seen for all the four sizes, but there is considerable finite size effect. For
the system sizes $L = 24$ and $L = 28$ the transition is quite sharp and
recognizably discontinuous to the eye, but further analysis discussed below
makes it clear that it is indeed of the first order. It can be
noted that as the view-angle is decreased, the order  parameter varies slowly
but non-monotonically. It at first decreases slightly, and then increases to a 
maximum near the transition, and then drops to low values (not exactly zero,
due to the finite size of the system) that characterize a disordered state. 

\begin{figure} [!h]
\includegraphics[trim=0cm 0cm 0cm 0cm, scale = 0.5]{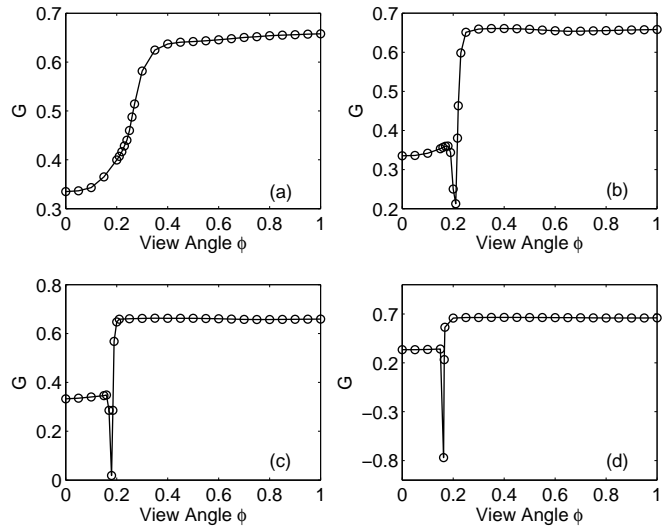}
\caption{\label{G_vs_phi}
Binder cumulant $G$ \emph{vs} view-angle $\phi$ (in units of $\pi$) plots for
the system sizes (a)
$L=16$ (b) $L=20$ (c) $L=24$ (d) $L=28$.
}
\end{figure}

In numerical simulations it is not a simple matter to estimate precisely the
critical value of the control parameter and to determine the order of the
phase-transition, because the finite size effects `round and shift' the
transition. A complete treatment of the numerical simulation of a phase
transition would require a full-fledged finite size scaling (FSS) analysis
\cite{binder, privman}. This theory of finite size scaling was originally
developed for the equilibrium phase transition, but now it is known that much of
this analysis is applicable to far-from-equilibrium phase transitions also, such
as the ones observed in the Vicsek model \cite{Lubeck2004, Henkel2008, chate,
baglietto2008}. Here we do not wish to carry out a full finite size
scaling analysis for the phase transitions we have observed, because that would
require very large system sizes and times, and is quite prohibitive for us at
present. Here we only wish to estimate the critical value of the control
parameter $\phi$ with moderate accuracy and establish that the phase transition
is of the first order. This can be done using a function of the moments of the
order parameter, known as the `Binder cumulant', defined by \cite{binder}
 \begin{equation}
 G(\eta,L) =1 - \frac{\langle\psi^4(t)\rangle} {3 \langle {\psi^2(t)} \rangle
^2},   \label{bindcum}
 \end{equation}
 where $\langle\psi^2(t)\rangle$ and $\langle\psi^4(t)\rangle$ are
time-averaged second and fourth moments of the order parameter, for given
values of $\eta$ and $L$ (for averaging  details please  see the previous
section). For second order phase transition the Binder cumulant is known to take
a value independent of the system size $L$, and so if the Binder cumulant is
plotted as a function of the control parameter for different  system sizes all
the curves are expected to cross at the critical value of the  order parameter.
On the other hand, for first order phase transitions the Binder cumulant
dips towards the negative values at the transition point, the dip becoming
sharper and deeper as the system size increases. In Fig. \ref{G_vs_phi} we have
shown the plots for Binder cumulant $G$ as a function of the view-angle  $\phi$.
For the system size $L = 16$ we do not see any dip, as the finite size  effects
are too large. But for system sizes $L = 20, 24, 28$ we have very clear dips.
Taking the value $\phi$  for the minimum of $G$ as the estimated critical
value $\phi_c$, we get $\phi_c = 0.21\pi, 0.18\pi, 0.1625\pi$ for the system
sizes $L = 20, 24, 28$ respectively. Also, we get $G \approx 2/3$ in the 
ordered phase and $G \approx 1/3$ in the disordered state~\cite{chate}, as 
expected.

\begin{figure} [!ht]
\includegraphics[trim=0cm 0cm 0cm 0cm, scale = 0.5]{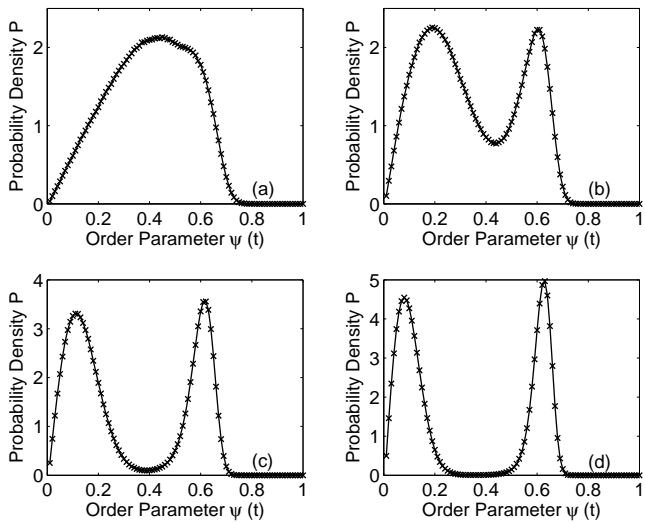}
\caption{\label{p_vs_phi}
Probability distribution of the instantaneous order parameter $\psi(t)$ for 
(a)$L=16$, $\phi$ = 0.270$\pi$ 
(b)$L=20$, $\phi$ = 0.217$\pi$ 
(c)$L=24$,  $\phi$ = 0.185$\pi$ 
(d) $L=28$, $\phi$ = 0.165$\pi$.
} 
\end{figure}

\begin{figure} [!ht]
\includegraphics[trim=0cm 0cm 0cm 0cm, scale = 0.5]{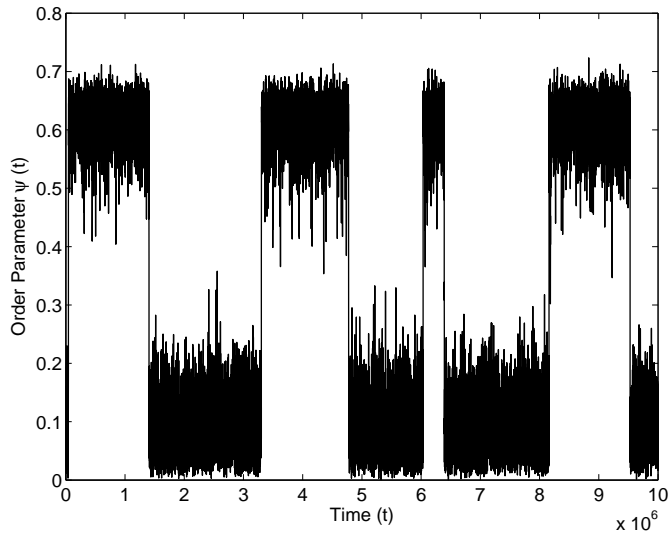}
\caption{\label{psi_vs_t}
A section of instantaneous order parameter $\psi(t)$ time-series close to the
phase transition. The jumps between ordered state (with $\langle \psi(t)
\rangle \approx 0.6$ ) and disordered state (with $\langle \psi(t)
\rangle \approx 0.1$ ) are clearly seen. Here the system size $L = 28$ and the
view-angle $\phi = 0.165\pi$.
} 
\end{figure}

We have also calculated the distribution of the instantaneous  order parameter
$\psi(t)$ in Fig. \ref{p_vs_phi} close to the estimated critical
points. For a  first order phase transition at the transition point both the
phases coexist, and over a period of time the system fluctuates between the
ordered and disordered states. The distribution for the system size $L =16$ 
(Fig. \ref{p_vs_phi}(a) )is a broad unimodal curve, which means due to the
finite size effects the distinction between two phases is blurred, consistent 
with the binder cumulant plot for the same system size (Fig. \ref{G_vs_phi}(a)).
 But for the system sizes $L = 20, 24, 28$ (Figs. \ref{p_vs_phi} (b), (c) and
(d) respectively) we have clear bimodal distributions, showing the presence of
both the ordered and disordered phases. This is also seen in Fig.
\ref{psi_vs_t}, which gives a sample of the time series $\psi(t)$ for the system
size $L = 28$. Here we clearly see the system abruptly and stochastically
switching between the two phases, one with average order parameter  $ \langle
\psi(t) \rangle \approx  0.6$ (the ordered  state) and the one with
average order parameter $ \langle \psi(t) \rangle \approx  0.1$ 
(the disordered state), which agrees with the distribution peak positions in
Fig.  \ref{p_vs_phi}.

\begin{figure} [!ht]
\includegraphics[trim=0cm 0cm 0cm 0cm, scale = 0.5]{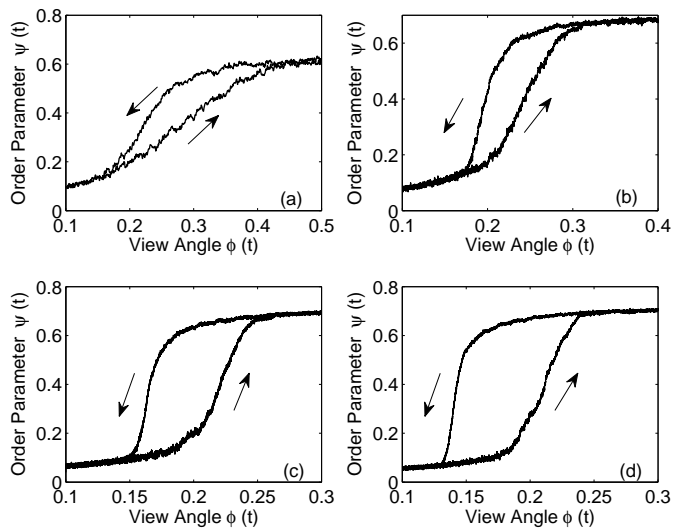}
\caption{\label{hysterisis}
Hysteresis in the variation of the instantaneous order parameter $\psi(t)$ with
the instantaneous view angle $\phi(t)$ (in units of $\pi$) for
(a)$L=16$, ramp rate $1.15 \times 10^{-4}$
(b)$L=20$, ramp rate $1.50 \times 10^{-5}$ 
(c)$L=24$, ramp rate $6.66 \times 10^{-6}$  
(d)$L=28$, ramp rate $6.66 \times 10^{-6}$. 
The ramp-rate is in radians/unit time. The arrows indicate ramp-up and
ramp-down sections of the hysteresis curves.  Each hysteresis loop is obtained
by averaging over $100$ independent realizations.
} 
\end{figure}

Another signature of a first order phase transition is the  presence of
hysteresis phenomenon near the transition point. If the control parameter is
ramped up and down across the critical point at a small, constant ramp-rate, the
instantaneous order parameter shows hysteresis. Fig. \ref{hysterisis} shows the
hysteresis in the instantaneous order parameter $\psi(t)$ as the view-angle
$\phi$ is ramped up and down at small ramp-rates in the range of $1.5\times
10^{-5}$ to $6.66\times 10^{-6}$ radians/unit time. We obtain the clear
hysteresis loops for all the four system sizes. The loops are centered about the
view-angle $\phi$ values which match the critical $\phi$ values as estimated
from Figs. \ref{psi_vs_phi} and \ref{G_vs_phi}.

\begin{figure} [!ht]
\includegraphics[trim=0cm 0cm 0cm 0cm, scale = 0.5]{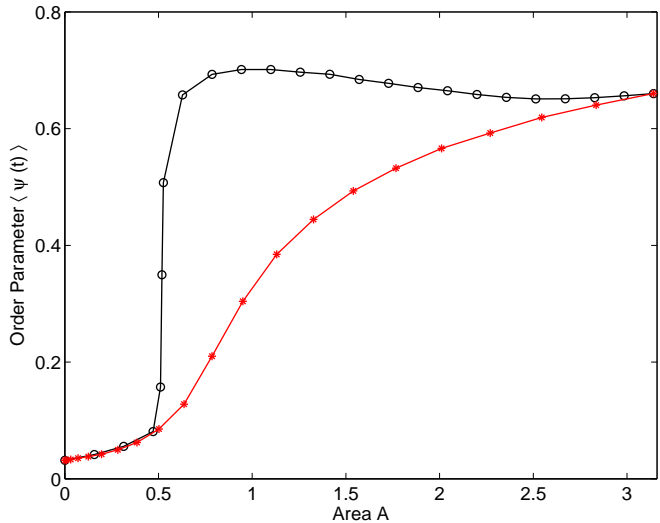}
\caption{\label{areatrans}
(Color online)
Order parameter $\langle \psi(t) \rangle$ as a function of the area of the
interaction neighborhood $A$. The black curve is for fixed radius $R=1$
neighborhood (for which $A = \phi$ numerically) and the red curve for the
fixed view-angle $\phi = \pi$ (for which $A = R^2$ numerically). Both curves
are for the system size $L = 28$. 
} 
\end{figure}

\begin{figure} [!ht]
\includegraphics[trim=0cm 0cm 0cm 0cm, scale = 0.5]{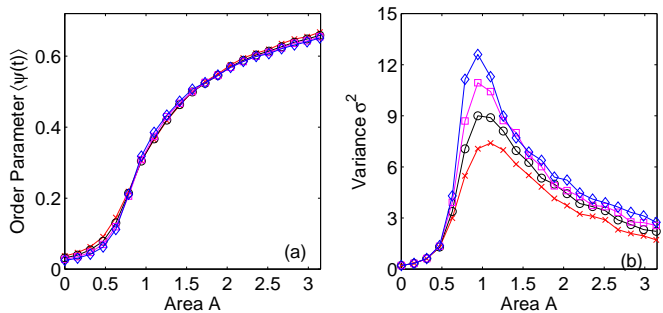}
\caption{\label{R_vary}
(Color online)
The plots of (a) order parameter $\langle \psi(t) \rangle$ \emph{vs} interaction
neighborhood area $A$(in units of $\pi$), (b) Variance $\sigma^2(A,L)$ 
\emph{vs} neighborhood area $A$ (in units of $\pi$). The area varied by varying 
the radius $R$, with view-angle $\phi = \pi$ fixed. so that numerically $A = 
R^2$. The symbols cross, circle, square and diamond respectively correspond to 
$L = 24, 28, 32, 36$ in that order. 
} 
\end{figure}

The above results show the effects of the variation of the view-angle on the
collective motion of the particles. As we reduce the view-angle we are in
effect doing two things -- we are reducing the size of the neighborhood, and we
are also introducing an increasing degree of anisotropy or directionality in the
interaction of the particle with its neighbors. Therefore it would be of
interest to know how the effects would differ if the size of the neighborhood is
varied while the interaction remains isotropic. In Fig. \ref{areatrans} we have
shown a comparison of the variation of the order parameter $\langle \psi(t)
\rangle$ as a function of the size of the neighborhood. The size, as measured
by the area of the neighborhood $A$, is varied in in two ways -- by varying the
view-angle $\phi$ from $0$ to $\pi$ (as radius $R=1$ is held constant), and by
varying the radius $R$ from $0$ to $1$ (as the view-angle $\phi = \pi$ is held
constant). The results differ qualitatively. In the first case, as we have
already discussed above, we have a first order phase transition at the
critical area $A_c \approx 0.5$ (which corresponds to $\phi \approx 0.16\pi$)
and a non-monotonic variation of the order parameter with a maximum around
$A \approx 1$. In the second case we have a monotonic fall in the order
parameter as the area of the neighborhood decreases, with no obvious
indication of a phase transition. But in fact there does appear to be a second
order phase transition which is obscured by the finite size effects. This can be
seen from the behavior of the variance of the order parameter $\sigma^2(A,L)$
as a function of the neighborhood area $A$, defined by \cite{chate,
baglietto2008}
\begin{equation}
 \sigma^2(A, L) = L^2[\langle\psi^2(t)\rangle-\langle\psi(t)\rangle^2].
\label{vararea}
\end{equation}
This variance is expected to diverge at a second order phase transition. In
Fig. \ref{R_vary} we have shown both the order parameter and the variance of
the order parameter as functions of the neighborhood area, which is varied by
varying the neighborhood radius holding the view-angle fixed at the maximum
value $\phi = \pi$. In Fig. \ref{R_vary} (a), which shows the order
parameter as a function of the neighborhood area $A$ for four different system
sizes $L = 24, 28, 32, 36$,  the phase-transition is not clearly discernible.
But in Fig. \ref{R_vary}(b), which shows the variance of the order parameter as
a function of the neighborhood area, we have a clear
peak around $A \approx 1$ which becomes more pronounced as the system size
increases (it is expected to diverge as $L \rightarrow \infty $). To establish
the presence of this phase-transition conclusively and to characterize its
nature (i.e., second or first order transition) one would need
to do detailed finite size scaling analysis. Here we only wish to underline the
point that the first-order phase transition with view-angle as the control
parameter discussed in this work is not only the effect of the variation in the
size of the interaction neighborhood, but also the effect of the change in the
degree of directionality of the interaction.

Before concluding, we shall discuss one recent report by Nguyen et 
al.\cite{nguyen}, who also have studied the same model as we have in this
study.  In their work they estimate the critical noise $\eta_c$ as a function of
the  view-angles $\phi$. Based on their observations they claim that for $\phi <
0.5\pi$ the critical noise is zero, or in other words the ordered state does not
exist for the view-angles $\phi < 0.5\pi$  in the presence of the noise (i.e,
$\eta > 0$). They discuss the implication of this observation to the presence
of the phenomenon of collective motion (which they call flocking) in certain
animal species and its absence in others. They draw the  conclusion that the
prey species, which presumably have view-angle $\phi > 0.5\pi$, do display
flocking behavior; whereas the predator species with view angle  $\phi < 0.5\pi$
do not. This claim would be more substantial if this critical view angle
value of $0.5\pi$ were robust to some variation in other parameters (velocity
$v$, density $\rho$, radius of interaction $R$); because the parameter values
they have chosen $\rho = 1.0, R = 1, v = 1.0$ have no special physical
significance. But as we have seen in the present study, where $\rho$ and $R$
have same values but $v = 0.5$, the ordered state does persist all the way down
to $\phi = 0.2\pi$ (for non-zero noise, i.e, $\eta = 0.3$). In fact, we obtain a
higher degree of order at $\phi = 0.35\pi$ than at $\phi = 1.0\pi$.

\section{Conclusions}

We have done a Monte Carlo simulation study of a modified Vicsek model in two 
dimensions. We have studied the effect of the variation of the view-angle on the
collective motion of the particles. We found that as the view-angle is reduced
the order in the system varies non-monotonically; the order parameter at first
decreases slowly and then increases, attaining a maximum at a remarkably low
view-angle, just before the system undergoes a first order phase
transition to a disordered state. The results are qualitatively different when
we reduce the radius of the (circular) neighborhood -- in this case the order
parameter decreases monotonically and goes to zero continuously. And the
variance of the order parameter shows a peak for a certain value of the
neighborhood radius, suggesting there could be a second order order-disorder
phase transition. Considering the importance of the limited view-angle in
modeling the motion of real world systems, such as flocks of birds, it would be
interesting to study if similar phase transitions arise in other models of
self-propelled particles, when the angular range of inter-particle interaction
in such models is varied.

\section{ACKNOWLEDGMENTS}

We acknowledge financial support from Board of College and University 
Development, Savitribai Phule Pune University. This work was carried with HPC
facilities provided by Centre for  Development of Advanced Computing (CDAC) and
also HPC facilities under the DST-FIST program  at the Department of Physics,
SPPU.

We wish to thank Prof. R. K. Pathak and Dr. P. Durganandini for helpful
discussions. 
\bibliography{main}

\begin{thebibliography}{37}%
\makeatletter
\providecommand \@ifxundefined [1]{%
 \@ifx{#1\undefined}
}%
\providecommand \@ifnum [1]{%
 \ifnum #1\expandafter \@firstoftwo
 \else \expandafter \@secondoftwo
 \fi
}%
\providecommand \@ifx [1]{%
 \ifx #1\expandafter \@firstoftwo
 \else \expandafter \@secondoftwo
 \fi
}%
\providecommand \natexlab [1]{#1}%
\providecommand \enquote  [1]{``#1''}%
\providecommand \bibnamefont  [1]{#1}%
\providecommand \bibfnamefont [1]{#1}%
\providecommand \citenamefont [1]{#1}%
\providecommand \href@noop [0]{\@secondoftwo}%
\providecommand \href [0]{\begingroup \@sanitize@url \@href}%
\providecommand \@href[1]{\@@startlink{#1}\@@href}%
\providecommand \@@href[1]{\endgroup#1\@@endlink}%
\providecommand \@sanitize@url [0]{\catcode `\\12\catcode `\$12\catcode
  `\&12\catcode `\#12\catcode `\^12\catcode `\_12\catcode `\%12\relax}%
\providecommand \@@startlink[1]{}%
\providecommand \@@endlink[0]{}%
\providecommand \url  [0]{\begingroup\@sanitize@url \@url }%
\providecommand \@url [1]{\endgroup\@href {#1}{\urlprefix }}%
\providecommand \urlprefix  [0]{URL }%
\providecommand \Eprint [0]{\href }%
\providecommand \doibase [0]{http://dx.doi.org/}%
\providecommand \selectlanguage [0]{\@gobble}%
\providecommand \bibinfo  [0]{\@secondoftwo}%
\providecommand \bibfield  [0]{\@secondoftwo}%
\providecommand \translation [1]{[#1]}%
\providecommand \BibitemOpen [0]{}%
\providecommand \bibitemStop [0]{}%
\providecommand \bibitemNoStop [0]{.\EOS\space}%
\providecommand \EOS [0]{\spacefactor3000\relax}%
\providecommand \BibitemShut  [1]{\csname bibitem#1\endcsname}%
\let\auto@bib@innerbib\@empty
\bibitem [{\citenamefont {Becco}\ \emph {et~al.}(2006)\citenamefont {Becco},
  \citenamefont {Vandewalle}, \citenamefont {Delcourt},\ and\ \citenamefont
  {Poncin}}]{becco}%
  \BibitemOpen
  \bibfield  {author} {\bibinfo {author} {\bibfnamefont {C.}~\bibnamefont
  {Becco}}, \bibinfo {author} {\bibfnamefont {N.}~\bibnamefont {Vandewalle}},
  \bibinfo {author} {\bibfnamefont {J.}~\bibnamefont {Delcourt}}, \ and\
  \bibinfo {author} {\bibfnamefont {P.}~\bibnamefont {Poncin}},\ }\href@noop {}
  {\bibfield  {journal} {\bibinfo  {journal} {Phys A}\ }\textbf {\bibinfo
  {volume} {367}},\ \bibinfo {pages} {487} (\bibinfo {year}
  {2006})}\BibitemShut {NoStop}%
\bibitem [{\citenamefont {Couzin}\ and\ \citenamefont {Krause}(2003)}]{couzin}%
  \BibitemOpen
  \bibfield  {author} {\bibinfo {author} {\bibfnamefont {I.~D.}\ \bibnamefont
  {Couzin}}\ and\ \bibinfo {author} {\bibfnamefont {J.}~\bibnamefont
  {Krause}},\ }\href@noop {} {\bibfield  {journal} {\bibinfo  {journal} {Adv
  Study Behav}\ }\textbf {\bibinfo {volume} {32}},\ \bibinfo {pages} {1}
  (\bibinfo {year} {2003})}\BibitemShut {NoStop}%
\bibitem [{\citenamefont {Ballerini}\ \emph {et~al.}(2008)\citenamefont
  {Ballerini}, \citenamefont {Cabibbo}, \citenamefont {Candelier},
  \citenamefont {Cavagna}, \citenamefont {Cisbani}, \citenamefont {Giardina},
  \citenamefont {Orlandi}, \citenamefont {Parisi}, \citenamefont {Procaccini},
  \citenamefont {Viale},\ and\ \citenamefont {Zdravkovic}}]{ballerini}%
  \BibitemOpen
  \bibfield  {author} {\bibinfo {author} {\bibfnamefont {M.}~\bibnamefont
  {Ballerini}}, \bibinfo {author} {\bibfnamefont {N.}~\bibnamefont {Cabibbo}},
  \bibinfo {author} {\bibfnamefont {R.}~\bibnamefont {Candelier}}, \bibinfo
  {author} {\bibfnamefont {A.}~\bibnamefont {Cavagna}}, \bibinfo {author}
  {\bibfnamefont {E.}~\bibnamefont {Cisbani}}, \bibinfo {author} {\bibfnamefont
  {I.}~\bibnamefont {Giardina}}, \bibinfo {author} {\bibfnamefont
  {A.}~\bibnamefont {Orlandi}}, \bibinfo {author} {\bibfnamefont
  {G.}~\bibnamefont {Parisi}}, \bibinfo {author} {\bibfnamefont
  {A.}~\bibnamefont {Procaccini}}, \bibinfo {author} {\bibfnamefont
  {M.}~\bibnamefont {Viale}}, \ and\ \bibinfo {author} {\bibfnamefont
  {V.}~\bibnamefont {Zdravkovic}},\ }\href@noop {} {\bibfield  {journal}
  {\bibinfo  {journal} {Anim. Behav.}\ }\textbf {\bibinfo {volume} {76}},\
  \bibinfo {pages} {201} (\bibinfo {year} {2008})}\BibitemShut {NoStop}%
\bibitem [{\citenamefont {Zhang}\ \emph {et~al.}(2009)\citenamefont {Zhang},
  \citenamefont {Be'er}, \citenamefont {Smith}, \citenamefont {Florin},\ and\
  \citenamefont {Swinney}}]{zhang}%
  \BibitemOpen
  \bibfield  {author} {\bibinfo {author} {\bibfnamefont {H.~P.}\ \bibnamefont
  {Zhang}}, \bibinfo {author} {\bibfnamefont {A.}~\bibnamefont {Be'er}},
  \bibinfo {author} {\bibfnamefont {R.~S.}\ \bibnamefont {Smith}}, \bibinfo
  {author} {\bibfnamefont {E.-L.}\ \bibnamefont {Florin}}, \ and\ \bibinfo
  {author} {\bibfnamefont {H.~L.}\ \bibnamefont {Swinney}},\ }\href@noop {}
  {\bibfield  {journal} {\bibinfo  {journal} {Euro Phys Lett}\ }\textbf
  {\bibinfo {volume} {87}},\ \bibinfo {pages} {48011} (\bibinfo {year}
  {2009})}\BibitemShut {NoStop}%
\bibitem [{\citenamefont {Darnton}\ \emph {et~al.}(2010)\citenamefont
  {Darnton}, \citenamefont {Tuner}, \citenamefont {Rojevsky},\ and\
  \citenamefont {Berg}}]{darton}%
  \BibitemOpen
  \bibfield  {author} {\bibinfo {author} {\bibfnamefont {N.~C.}\ \bibnamefont
  {Darnton}}, \bibinfo {author} {\bibfnamefont {L.}~\bibnamefont {Tuner}},
  \bibinfo {author} {\bibfnamefont {S.}~\bibnamefont {Rojevsky}}, \ and\
  \bibinfo {author} {\bibfnamefont {H.~C.}\ \bibnamefont {Berg}},\ }\href@noop
  {} {\bibfield  {journal} {\bibinfo  {journal} {Biophys. J}\ }\textbf
  {\bibinfo {volume} {98}},\ \bibinfo {pages} {2082} (\bibinfo {year}
  {2010})}\BibitemShut {NoStop}%
\bibitem [{\citenamefont {Peruani}\ \emph {et~al.}(2012)\citenamefont
  {Peruani}, \citenamefont {Starru\ss}, \citenamefont {Jakovljevic},
  \citenamefont {S{\o}gaard-Anderson}, \citenamefont {Deutsch},\ and\
  \citenamefont {B{\"a}r}}]{peruani}%
  \BibitemOpen
  \bibfield  {author} {\bibinfo {author} {\bibfnamefont {F.}~\bibnamefont
  {Peruani}}, \bibinfo {author} {\bibfnamefont {J.}~\bibnamefont {Starru\ss}},
  \bibinfo {author} {\bibfnamefont {V.}~\bibnamefont {Jakovljevic}}, \bibinfo
  {author} {\bibfnamefont {L.}~\bibnamefont {S{\o}gaard-Anderson}}, \bibinfo
  {author} {\bibfnamefont {A.}~\bibnamefont {Deutsch}}, \ and\ \bibinfo
  {author} {\bibfnamefont {M.}~\bibnamefont {B{\"a}r}},\ }\href@noop {}
  {\bibfield  {journal} {\bibinfo  {journal} {Phy Rev Lett}\ }\textbf {\bibinfo
  {volume} {108}},\ \bibinfo {pages} {098102} (\bibinfo {year}
  {2012})}\BibitemShut {NoStop}%
\bibitem [{\citenamefont {Turgut}\ \emph {et~al.}(2008)\citenamefont {Turgut},
  \citenamefont {{\c C}elikkanat}, \citenamefont {G{\"o}k{\c c}e},\ and\
  \citenamefont {{\c S}ahin}}]{turgut}%
  \BibitemOpen
  \bibfield  {author} {\bibinfo {author} {\bibfnamefont {A.~E.}\ \bibnamefont
  {Turgut}}, \bibinfo {author} {\bibfnamefont {H.}~\bibnamefont {{\c
  C}elikkanat}}, \bibinfo {author} {\bibfnamefont {F.}~\bibnamefont {G{\"o}k{\c
  c}e}}, \ and\ \bibinfo {author} {\bibfnamefont {E.}~\bibnamefont {{\c
  S}ahin}},\ }\href@noop {} {\bibfield  {journal} {\bibinfo  {journal} {Swarm
  Intell}\ }\textbf {\bibinfo {volume} {2}},\ \bibinfo {pages} {97} (\bibinfo
  {year} {2008})}\BibitemShut {NoStop}%
\bibitem [{\citenamefont {Ke}\ \emph {et~al.}(2010)\citenamefont {Ke},
  \citenamefont {Ye}, \citenamefont {Carroll},\ and\ \citenamefont
  {Showalter}}]{ke}%
  \BibitemOpen
  \bibfield  {author} {\bibinfo {author} {\bibfnamefont {H.}~\bibnamefont
  {Ke}}, \bibinfo {author} {\bibfnamefont {S.}~\bibnamefont {Ye}}, \bibinfo
  {author} {\bibfnamefont {R.~L.}\ \bibnamefont {Carroll}}, \ and\ \bibinfo
  {author} {\bibfnamefont {K.}~\bibnamefont {Showalter}},\ }\href@noop {}
  {\bibfield  {journal} {\bibinfo  {journal} {J. Phys. Chem. A}\ }\textbf
  {\bibinfo {volume} {114}},\ \bibinfo {pages} {5462} (\bibinfo {year}
  {2010})}\BibitemShut {NoStop}%
\bibitem [{\citenamefont {Buttinoni}\ \emph {et~al.}(2013)\citenamefont
  {Buttinoni}, \citenamefont {Bialk{\'e}}, \citenamefont {K{\"u}mmel},
  \citenamefont {L{\"o}wen}, \citenamefont {Bechinger},\ and\ \citenamefont
  {Speck}}]{buttinoni}%
  \BibitemOpen
  \bibfield  {author} {\bibinfo {author} {\bibfnamefont {I.}~\bibnamefont
  {Buttinoni}}, \bibinfo {author} {\bibfnamefont {J.}~\bibnamefont
  {Bialk{\'e}}}, \bibinfo {author} {\bibfnamefont {F.}~\bibnamefont
  {K{\"u}mmel}}, \bibinfo {author} {\bibfnamefont {H.}~\bibnamefont
  {L{\"o}wen}}, \bibinfo {author} {\bibfnamefont {C.}~\bibnamefont
  {Bechinger}}, \ and\ \bibinfo {author} {\bibfnamefont {T.}~\bibnamefont
  {Speck}},\ }\href@noop {} {\bibfield  {journal} {\bibinfo  {journal} {Phys\
  Rev\ Lett}\ }\textbf {\bibinfo {volume} {110}},\ \bibinfo {pages} {238301}
  (\bibinfo {year} {2013})}\BibitemShut {NoStop}%
\bibitem [{\citenamefont {Kudrolli}\ \emph {et~al.}(2008)\citenamefont
  {Kudrolli}, \citenamefont {Lumay}, \citenamefont {Volfson},\ and\
  \citenamefont {Tsimring}}]{kudrolli}%
  \BibitemOpen
  \bibfield  {author} {\bibinfo {author} {\bibfnamefont {A.}~\bibnamefont
  {Kudrolli}}, \bibinfo {author} {\bibfnamefont {G.}~\bibnamefont {Lumay}},
  \bibinfo {author} {\bibfnamefont {D.}~\bibnamefont {Volfson}}, \ and\
  \bibinfo {author} {\bibfnamefont {L.~S.}\ \bibnamefont {Tsimring}},\
  }\href@noop {} {\bibfield  {journal} {\bibinfo  {journal} {Phys Rev Lett}\
  }\textbf {\bibinfo {volume} {100}},\ \bibinfo {pages} {058001} (\bibinfo
  {year} {2008})}\BibitemShut {NoStop}%
\bibitem [{\citenamefont {Aranson}\ \emph {et~al.}(2007)\citenamefont
  {Aranson}, \citenamefont {Volfson},\ and\ \citenamefont
  {Tsimring}}]{aranson}%
  \BibitemOpen
  \bibfield  {author} {\bibinfo {author} {\bibfnamefont {I.~S.}\ \bibnamefont
  {Aranson}}, \bibinfo {author} {\bibfnamefont {D.}~\bibnamefont {Volfson}}, \
  and\ \bibinfo {author} {\bibfnamefont {L.~S.}\ \bibnamefont {Tsimring}},\
  }\href@noop {} {\bibfield  {journal} {\bibinfo  {journal} {Phys Rev E}\
  }\textbf {\bibinfo {volume} {75}},\ \bibinfo {pages} {051301} (\bibinfo
  {year} {2007})}\BibitemShut {NoStop}%
\bibitem [{\citenamefont {Narayan}\ \emph {et~al.}(2006)\citenamefont
  {Narayan}, \citenamefont {Menon},\ and\ \citenamefont {Ramaswamy}}]{narayan}%
  \BibitemOpen
  \bibfield  {author} {\bibinfo {author} {\bibfnamefont {V.}~\bibnamefont
  {Narayan}}, \bibinfo {author} {\bibfnamefont {N.}~\bibnamefont {Menon}}, \
  and\ \bibinfo {author} {\bibfnamefont {S.}~\bibnamefont {Ramaswamy}},\
  }\href@noop {} {\bibfield  {journal} {\bibinfo  {journal} {J. Stat. Mech.}\
  }\textbf {\bibinfo {volume} {2006}},\ \bibinfo {pages} {P01005} (\bibinfo
  {year} {2006})}\BibitemShut {NoStop}%
\bibitem [{\citenamefont {Vicsek}\ \emph {et~al.}(1995)\citenamefont {Vicsek},
  \citenamefont {Czir{\'o}k}, \citenamefont {Ben-Jacob}, \citenamefont
  {Cohen},\ and\ \citenamefont {Shochet}}]{vicsek}%
  \BibitemOpen
  \bibfield  {author} {\bibinfo {author} {\bibfnamefont {T.}~\bibnamefont
  {Vicsek}}, \bibinfo {author} {\bibfnamefont {A.}~\bibnamefont {Czir{\'o}k}},
  \bibinfo {author} {\bibfnamefont {E.}~\bibnamefont {Ben-Jacob}}, \bibinfo
  {author} {\bibfnamefont {I.}~\bibnamefont {Cohen}}, \ and\ \bibinfo {author}
  {\bibfnamefont {O.}~\bibnamefont {Shochet}},\ }\href@noop {} {\bibfield
  {journal} {\bibinfo  {journal} {Phys.\ Rev.\ Lett}\ }\textbf {\bibinfo
  {volume} {75}},\ \bibinfo {pages} {1226} (\bibinfo {year}
  {1995})}\BibitemShut {NoStop}%
\bibitem [{\citenamefont {Chat{\'e}}\ \emph
  {et~al.}(2008{\natexlab{a}})\citenamefont {Chat{\'e}}, \citenamefont
  {Ginelli}, \citenamefont {Gr{\'e}goire},\ and\ \citenamefont
  {Raynaud}}]{chate}%
  \BibitemOpen
  \bibfield  {author} {\bibinfo {author} {\bibfnamefont {H.}~\bibnamefont
  {Chat{\'e}}}, \bibinfo {author} {\bibfnamefont {F.}~\bibnamefont {Ginelli}},
  \bibinfo {author} {\bibfnamefont {G.}~\bibnamefont {Gr{\'e}goire}}, \ and\
  \bibinfo {author} {\bibfnamefont {F.}~\bibnamefont {Raynaud}},\ }\href@noop
  {} {\bibfield  {journal} {\bibinfo  {journal} {Phys.\ Rev.\ E}\ }\textbf
  {\bibinfo {volume} {77}},\ \bibinfo {pages} {046113} (\bibinfo {year}
  {2008}{\natexlab{a}})}\BibitemShut {NoStop}%
\bibitem [{\citenamefont {Gr{\'e}goire}\ \emph {et~al.}(2003)\citenamefont
  {Gr{\'e}goire}, \citenamefont {Chat{\'e}},\ and\ \citenamefont {Tu}}]{tu}%
  \BibitemOpen
  \bibfield  {author} {\bibinfo {author} {\bibfnamefont {G.}~\bibnamefont
  {Gr{\'e}goire}}, \bibinfo {author} {\bibfnamefont {H.}~\bibnamefont
  {Chat{\'e}}}, \ and\ \bibinfo {author} {\bibfnamefont {Y.}~\bibnamefont
  {Tu}},\ }\href@noop {} {\bibfield  {journal} {\bibinfo  {journal} {Phys. D}\
  }\textbf {\bibinfo {volume} {181}},\ \bibinfo {pages} {157} (\bibinfo {year}
  {2003})}\BibitemShut {NoStop}%
\bibitem [{\citenamefont {Toner}\ and\ \citenamefont {Tu}(1998)}]{toner}%
  \BibitemOpen
  \bibfield  {author} {\bibinfo {author} {\bibfnamefont {J.}~\bibnamefont
  {Toner}}\ and\ \bibinfo {author} {\bibfnamefont {Y.}~\bibnamefont {Tu}},\
  }\href@noop {} {\bibfield  {journal} {\bibinfo  {journal} {Phys\ Rev\ E}\
  }\textbf {\bibinfo {volume} {58}},\ \bibinfo {pages} {4828} (\bibinfo {year}
  {1998})}\BibitemShut {NoStop}%
\bibitem [{\citenamefont {Levine}\ \emph {et~al.}(2000)\citenamefont {Levine},
  \citenamefont {Rappel},\ and\ \citenamefont {Cohen}}]{levin}%
  \BibitemOpen
  \bibfield  {author} {\bibinfo {author} {\bibfnamefont {H.}~\bibnamefont
  {Levine}}, \bibinfo {author} {\bibfnamefont {W.-J.}\ \bibnamefont {Rappel}},
  \ and\ \bibinfo {author} {\bibfnamefont {I.}~\bibnamefont {Cohen}},\
  }\href@noop {} {\bibfield  {journal} {\bibinfo  {journal} {Phys\ Rev\ E}\
  }\textbf {\bibinfo {volume} {63}},\ \bibinfo {pages} {017101} (\bibinfo
  {year} {2000})}\BibitemShut {NoStop}%
\bibitem [{\citenamefont {Gr{\'e}goire}\ and\ \citenamefont
  {Chat{\'e}}(2004)}]{gregoire}%
  \BibitemOpen
  \bibfield  {author} {\bibinfo {author} {\bibfnamefont {G.}~\bibnamefont
  {Gr{\'e}goire}}\ and\ \bibinfo {author} {\bibfnamefont {H.}~\bibnamefont
  {Chat{\'e}}},\ }\href@noop {} {\bibfield  {journal} {\bibinfo  {journal}
  {Phys\ Rev\ Lett}\ }\textbf {\bibinfo {volume} {92}},\ \bibinfo {pages}
  {025702} (\bibinfo {year} {2004})}\BibitemShut {NoStop}%
\bibitem [{\citenamefont {Chat{\'e}}\ \emph
  {et~al.}(2008{\natexlab{b}})\citenamefont {Chat{\'e}}, \citenamefont
  {Ginelli}, \citenamefont {Gregoir{\'e}}, \citenamefont {Peruani},\ and\
  \citenamefont {Raynaud}}]{ginelli}%
  \BibitemOpen
  \bibfield  {author} {\bibinfo {author} {\bibfnamefont {H.}~\bibnamefont
  {Chat{\'e}}}, \bibinfo {author} {\bibfnamefont {F.}~\bibnamefont {Ginelli}},
  \bibinfo {author} {\bibfnamefont {G.}~\bibnamefont {Gregoir{\'e}}}, \bibinfo
  {author} {\bibfnamefont {F.}~\bibnamefont {Peruani}}, \ and\ \bibinfo
  {author} {\bibfnamefont {F.}~\bibnamefont {Raynaud}},\ }\href@noop {}
  {\bibfield  {journal} {\bibinfo  {journal} {Eur. Phys. J. B.}\ }\textbf
  {\bibinfo {volume} {64}},\ \bibinfo {pages} {451} (\bibinfo {year}
  {2008}{\natexlab{b}})}\BibitemShut {NoStop}%
\bibitem [{\citenamefont {Vicsek}\ and\ \citenamefont
  {Zafeiris}(2012)}]{zafeiris}%
  \BibitemOpen
  \bibfield  {author} {\bibinfo {author} {\bibfnamefont {T.}~\bibnamefont
  {Vicsek}}\ and\ \bibinfo {author} {\bibfnamefont {A.}~\bibnamefont
  {Zafeiris}},\ }\href@noop {} {\bibfield  {journal} {\bibinfo  {journal}
  {Phys. Rep.}\ }\textbf {\bibinfo {volume} {517}},\ \bibinfo {pages} {71}
  (\bibinfo {year} {2012})}\BibitemShut {NoStop}%
\bibitem [{\citenamefont {Menzel}(2015)}]{menzel}%
  \BibitemOpen
  \bibfield  {author} {\bibinfo {author} {\bibfnamefont {A.}~\bibnamefont
  {Menzel}},\ }\href@noop {} {\bibfield  {journal} {\bibinfo  {journal} {Phys.
  Rep.}\ }\textbf {\bibinfo {volume} {554}},\ \bibinfo {pages} {1} (\bibinfo
  {year} {2015})}\BibitemShut {NoStop}%
\bibitem [{\citenamefont {Tian}\ \emph {et~al.}(2009)\citenamefont {Tian},
  \citenamefont {Yang}, \citenamefont {Li}, \citenamefont {Wang}, \citenamefont
  {Wang},\ and\ \citenamefont {Zhou}}]{tian}%
  \BibitemOpen
  \bibfield  {author} {\bibinfo {author} {\bibfnamefont {B.-M.}\ \bibnamefont
  {Tian}}, \bibinfo {author} {\bibfnamefont {H.-X.}\ \bibnamefont {Yang}},
  \bibinfo {author} {\bibfnamefont {W.}~\bibnamefont {Li}}, \bibinfo {author}
  {\bibfnamefont {W.-X.}\ \bibnamefont {Wang}}, \bibinfo {author}
  {\bibfnamefont {B.-H.}\ \bibnamefont {Wang}}, \ and\ \bibinfo {author}
  {\bibfnamefont {T.}~\bibnamefont {Zhou}},\ }\href@noop {} {\bibfield
  {journal} {\bibinfo  {journal} {Phys.\ Rev.\ E}\ }\textbf {\bibinfo {volume}
  {79}},\ \bibinfo {pages} {052102} (\bibinfo {year} {2009})}\BibitemShut
  {NoStop}%
\bibitem [{\citenamefont {Li}\ \emph {et~al.}(2011)\citenamefont {Li},
  \citenamefont {Wang}, \citenamefont {Han}, \citenamefont {Tian},
  \citenamefont {Xi},\ and\ \citenamefont {Wang}}]{li}%
  \BibitemOpen
  \bibfield  {author} {\bibinfo {author} {\bibfnamefont {Y.-J.}\ \bibnamefont
  {Li}}, \bibinfo {author} {\bibfnamefont {S.}~\bibnamefont {Wang}}, \bibinfo
  {author} {\bibfnamefont {Z.-L.}\ \bibnamefont {Han}}, \bibinfo {author}
  {\bibfnamefont {B.-M.}\ \bibnamefont {Tian}}, \bibinfo {author}
  {\bibfnamefont {Z.-D.}\ \bibnamefont {Xi}}, \ and\ \bibinfo {author}
  {\bibfnamefont {B.-H.}\ \bibnamefont {Wang}},\ }\href@noop {} {\bibfield
  {journal} {\bibinfo  {journal} {Eur. Phys. J.}\ }\textbf {\bibinfo {volume}
  {93}},\ \bibinfo {pages} {68003} (\bibinfo {year} {2011})}\BibitemShut
  {NoStop}%
\bibitem [{\citenamefont {Gao}\ \emph {et~al.}(2011)\citenamefont {Gao},
  \citenamefont {Havlin}, \citenamefont {Xu},\ and\ \citenamefont
  {Stanley}}]{gao}%
  \BibitemOpen
  \bibfield  {author} {\bibinfo {author} {\bibfnamefont {J.}~\bibnamefont
  {Gao}}, \bibinfo {author} {\bibfnamefont {S.}~\bibnamefont {Havlin}},
  \bibinfo {author} {\bibfnamefont {X.}~\bibnamefont {Xu}}, \ and\ \bibinfo
  {author} {\bibfnamefont {H.~E.}\ \bibnamefont {Stanley}},\ }\href@noop {}
  {\bibfield  {journal} {\bibinfo  {journal} {Phys\ Rev\ E}\ }\textbf {\bibinfo
  {volume} {84}},\ \bibinfo {pages} {046115} (\bibinfo {year}
  {2011})}\BibitemShut {NoStop}%
\bibitem [{\citenamefont {Yang}\ and\ \citenamefont {Rong}(2015)}]{yang}%
  \BibitemOpen
  \bibfield  {author} {\bibinfo {author} {\bibfnamefont {H.-X.}\ \bibnamefont
  {Yang}}\ and\ \bibinfo {author} {\bibfnamefont {Z.}~\bibnamefont {Rong}},\
  }\href@noop {} {\bibfield  {journal} {\bibinfo  {journal} {Phys. A}\ }\textbf
  {\bibinfo {volume} {432}},\ \bibinfo {pages} {180} (\bibinfo {year}
  {2015})}\BibitemShut {NoStop}%
\bibitem [{\citenamefont {McComb}\ and\ \citenamefont
  {Kajiura}(2008)}]{mccomb}%
  \BibitemOpen
  \bibfield  {author} {\bibinfo {author} {\bibfnamefont {D.}~\bibnamefont
  {McComb}}\ and\ \bibinfo {author} {\bibfnamefont {S.}~\bibnamefont
  {Kajiura}},\ }\href@noop {} {\bibfield  {journal} {\bibinfo  {journal} {J.
  Exp. Biol.}\ }\textbf {\bibinfo {volume} {211}},\ \bibinfo {pages} {482}
  (\bibinfo {year} {2008})}\BibitemShut {NoStop}%
\bibitem [{\citenamefont {Martin}\ and\ \citenamefont {Katzir}(1999)}]{martin}%
  \BibitemOpen
  \bibfield  {author} {\bibinfo {author} {\bibfnamefont {G.}~\bibnamefont
  {Martin}}\ and\ \bibinfo {author} {\bibfnamefont {G.}~\bibnamefont
  {Katzir}},\ }\href@noop {} {\bibfield  {journal} {\bibinfo  {journal} {Brain
  Behav Evol}\ }\textbf {\bibinfo {volume} {53}},\ \bibinfo {pages} {55}
  (\bibinfo {year} {1999})}\BibitemShut {NoStop}%
\bibitem [{\citenamefont {J{\'e}kely}\ \emph {et~al.}(2008)\citenamefont
  {J{\'e}kely}, \citenamefont {Colombbelli}, \citenamefont {Hausen},
  \citenamefont {Guy}, \citenamefont {Stelzer}, \citenamefont
  {N{\'e}d{\'e}lec},\ and\ \citenamefont {Arendt}}]{jekely}%
  \BibitemOpen
  \bibfield  {author} {\bibinfo {author} {\bibfnamefont {G.}~\bibnamefont
  {J{\'e}kely}}, \bibinfo {author} {\bibfnamefont {J.}~\bibnamefont
  {Colombbelli}}, \bibinfo {author} {\bibfnamefont {H.}~\bibnamefont {Hausen}},
  \bibinfo {author} {\bibfnamefont {K.}~\bibnamefont {Guy}}, \bibinfo {author}
  {\bibfnamefont {E.}~\bibnamefont {Stelzer}}, \bibinfo {author} {\bibfnamefont
  {F.}~\bibnamefont {N{\'e}d{\'e}lec}}, \ and\ \bibinfo {author} {\bibfnamefont
  {D.}~\bibnamefont {Arendt}},\ }\href@noop {} {\bibfield  {journal} {\bibinfo
  {journal} {Nature}\ }\textbf {\bibinfo {volume} {456}},\ \bibinfo {pages}
  {395} (\bibinfo {year} {2008})}\BibitemShut {NoStop}%
\bibitem [{\citenamefont {Baglietto}\ and\ \citenamefont
  {Albano}(2009)}]{baglietto}%
  \BibitemOpen
  \bibfield  {author} {\bibinfo {author} {\bibfnamefont {G.}~\bibnamefont
  {Baglietto}}\ and\ \bibinfo {author} {\bibfnamefont {E.~V.}\ \bibnamefont
  {Albano}},\ }\href@noop {} {\bibfield  {journal} {\bibinfo  {journal} {Phys\
  Rev\ E}\ }\textbf {\bibinfo {volume} {80}},\ \bibinfo {pages} {050103(R)}
  (\bibinfo {year} {2009})}\BibitemShut {NoStop}%
\bibitem [{\citenamefont {Gould}\ \emph {et~al.}(2006)\citenamefont {Gould},
  \citenamefont {Tobochnik},\ and\ \citenamefont {Christian}}]{gould}%
  \BibitemOpen
  \bibfield  {author} {\bibinfo {author} {\bibfnamefont {H.}~\bibnamefont
  {Gould}}, \bibinfo {author} {\bibfnamefont {J.}~\bibnamefont {Tobochnik}}, \
  and\ \bibinfo {author} {\bibfnamefont {W.}~\bibnamefont {Christian}},\
  }\href@noop {} {\emph {\bibinfo {title} {An Introduction to Computer
  Simulationn Methods: Applications to Physical Systems}}}\ (\bibinfo
  {publisher} {Addison-Wesley},\ \bibinfo {address} {San Fransisco},\ \bibinfo
  {year} {2006})\BibitemShut {NoStop}%
\bibitem [{\citenamefont {Wang}\ \emph {et~al.}(2013)\citenamefont {Wang},
  \citenamefont {Zhu}, \citenamefont {Yin}, \citenamefont {Hu},\ and\
  \citenamefont {Yan}}]{wang}%
  \BibitemOpen
  \bibfield  {author} {\bibinfo {author} {\bibfnamefont {X.-G.}\ \bibnamefont
  {Wang}}, \bibinfo {author} {\bibfnamefont {C.-P.}\ \bibnamefont {Zhu}},
  \bibinfo {author} {\bibfnamefont {C.-Y.}\ \bibnamefont {Yin}}, \bibinfo
  {author} {\bibfnamefont {D.-S.}\ \bibnamefont {Hu}}, \ and\ \bibinfo {author}
  {\bibfnamefont {Z.-J.}\ \bibnamefont {Yan}},\ }\href@noop {} {\bibfield
  {journal} {\bibinfo  {journal} {Phys. A}\ }\textbf {\bibinfo {volume}
  {392}},\ \bibinfo {pages} {2398} (\bibinfo {year} {2013})}\BibitemShut
  {NoStop}%
\bibitem [{\citenamefont {Binder}(1997)}]{binder}%
  \BibitemOpen
  \bibfield  {author} {\bibinfo {author} {\bibfnamefont {K.}~\bibnamefont
  {Binder}},\ }\href@noop {} {\bibfield  {journal} {\bibinfo  {journal} {Rep.
  Prog. Phys.}\ }\textbf {\bibinfo {volume} {60}},\ \bibinfo {pages} {487}
  (\bibinfo {year} {1997})}\BibitemShut {NoStop}%
\bibitem [{\citenamefont {Privman}(1990)}]{privman}%
  \BibitemOpen
  \bibfield  {author} {\bibinfo {author} {\bibfnamefont {E.~B.}\ \bibnamefont
  {Privman}},\ }\href@noop {} {\emph {\bibinfo {title} {Finite Size Scaling and
  Numerical Simulations of Statistical Systems}}}\ (\bibinfo  {publisher}
  {World Scientific Publishing},\ \bibinfo {address} {Singapore},\ \bibinfo
  {year} {1990})\BibitemShut {NoStop}%
\bibitem [{\citenamefont {S.L\"ubeck}(2004)}]{Lubeck2004}%
  \BibitemOpen
  \bibfield  {author} {\bibinfo {author} {\bibnamefont {S.L\"ubeck}},\
  }\href@noop {} {\bibfield  {journal} {\bibinfo  {journal} {Int. \ J. \ Mod. \
  Phys. \ B}\ }\textbf {\bibinfo {volume} {18}},\ \bibinfo {pages} {3977}
  (\bibinfo {year} {2004})}\BibitemShut {NoStop}%
\bibitem [{\citenamefont {Henkel}\ \emph {et~al.}(2008)\citenamefont {Henkel},
  \citenamefont {Hinrichsen},\ and\ \citenamefont {Lubeck}}]{Henkel2008}%
  \BibitemOpen
  \bibfield  {author} {\bibinfo {author} {\bibfnamefont {M.}~\bibnamefont
  {Henkel}}, \bibinfo {author} {\bibfnamefont {H.}~\bibnamefont {Hinrichsen}},
  \ and\ \bibinfo {author} {\bibfnamefont {S.}~\bibnamefont {Lubeck}},\
  }\href@noop {} {\emph {\bibinfo {title} {Non-equilibrium Phase Transitions,
  Volume 1}}}\ (\bibinfo  {publisher} {Springer},\ \bibinfo {year}
  {2008})\BibitemShut {NoStop}%
\bibitem [{\citenamefont {Baglietto}\ and\ \citenamefont
  {Albano}(2008)}]{baglietto2008}%
  \BibitemOpen
  \bibfield  {author} {\bibinfo {author} {\bibfnamefont {G.}~\bibnamefont
  {Baglietto}}\ and\ \bibinfo {author} {\bibfnamefont {E.~V.}\ \bibnamefont
  {Albano}},\ }\href@noop {} {\bibfield  {journal} {\bibinfo  {journal} {Phys\
  Rev\ E}\ }\textbf {\bibinfo {volume} {78}},\ \bibinfo {pages} {021125}
  (\bibinfo {year} {2008})}\BibitemShut {NoStop}%
\bibitem [{\citenamefont {Nguyen}\ \emph {et~al.}(2015)\citenamefont {Nguyen},
  \citenamefont {Lee},\ and\ \citenamefont {Ngo}}]{nguyen}%
  \BibitemOpen
  \bibfield  {author} {\bibinfo {author} {\bibfnamefont {P.~T.}\ \bibnamefont
  {Nguyen}}, \bibinfo {author} {\bibfnamefont {S.-H.}\ \bibnamefont {Lee}}, \
  and\ \bibinfo {author} {\bibfnamefont {V.~T.}\ \bibnamefont {Ngo}},\
  }\href@noop {} {\bibfield  {journal} {\bibinfo  {journal} {Phys\ Rev\ E}\
  }\textbf {\bibinfo {volume} {92}},\ \bibinfo {pages} {032716} (\bibinfo
  {year} {2015})}\BibitemShut {NoStop}%
\end{thebibliography}%

\end{document}